\documentclass[12pt,preprint]{aastex}

\begin{document}

\shorttitle{A new channel for type Ia supernovae}

\shortauthors{Rosswog, Kasen, Guillochon \& Ramirez-Ruiz}

\title{Collisions of white dwarfs as a new progenitor channel for type  
Ia supernovae}

\author{Stephan Rosswog\altaffilmark{1}, Daniel Kasen\altaffilmark{2,3}, James Guillochon\altaffilmark{2}, and Enrico Ramirez-Ruiz\altaffilmark{2}}
\altaffiltext{1}{School of Engineering and Science, Jacobs University
  Bremen, Campus Ring 1, 28759 Bremen, Germany; s.rosswog@jacobs-university.de}
\altaffiltext{2}{Department of Astronomy and Astrophysics, University
  of California, Santa Cruz, CA 95064; enrico@ucolick.org}
 \altaffiltext{3}{Hubble Fellow}
  
\newcommand{\Nifs}{\ensuremath{^{56}\mathrm{Ni}}}
\def\paren#1{\left( #1 \right)}
\def\Mesz{M\'esz\'aros~}
\def\Pacz{Paczy\'nski~}
\def\Kluz{Klu\'zniak~}
\def\p{$e^\pm \;$}
\def\msun{M$_{\odot}$}
\def\Msun{M$_{\odot}$ }
\def\be{\begin{equation}}
\def\ee{\end{equation}}
\def\bi{\begin{itemize}}
\def\ei{\end{itemize}}
\def\bea{\begin{eqnarray}}
\def\eea{\end{eqnarray}}
\def\gcc{gcm$^{-3}$ }
\def\edo{\end{document}}
\def\red{\textcolor{red}}
\def\blue{\textcolor{blue}}
\def\green{\textcolor{green}}

\begin{abstract}
We present the results of  a systematic numerical study of an  
alternative progenitor scenario to produce type Ia supernova  
explosions, which is not restricted to the ignition of a  
CO white dwarf near the Chandrasekhar mass. In this scenario,  
a shock-triggered thermonuclear  
explosion ensues from the collision of two white dwarfs. Consistent  
modeling of the gas dynamics together with nuclear reactions using   
both a smoothed particle and a grid-based  hydrodynamics code are  
performed to study the viability of  this alternative progenitor  
channel.  
We find that shock-triggered  
ignition and the synthesis of Ni are in fact a natural outcome for  
moderately massive white dwarf pairs colliding close to head-on.  
We use a multi-dimensional radiative transfer code to calculate the  
emergent broadband light curves and spectral time-series of these  
events. The synthetic spectra and lightcurves compare well with those  
of normal type Ia supernovae  over a similar B-band decline rate and   
are broadly consistent with the Phillips relation, although a mild  
dependence on viewing angle is observed due to the asymmetry of the  
ejected debris. While event rates within galactic centers and globular  
clusters are found to be much too low to explain the bulk of the type  
Ia supernovae population, they may be frequent enough to make as much  
as a one percent contribution to the overall rate.  Although these  
rate estimates are still subject to substantial uncertainties, they do  
suggest that  dense stellar systems should provide upcoming supernova  
surveys with hundreds of  such collision-induced thermonuclear  
explosions  per year.
\end{abstract}


\section{Introduction}
Type Ia supernovae (SNe Ia) are of major astrophysical relevance. They  
have acquired particular cosmological significance as a 
probe of the scale and geometry of the universe,  
providing the first evidence for its acceleration  
\citep{riess98,perlmutter99,astier06}. These results depend crucially  
on the assumption that SNe Ia are standard candles. This assumption  
could be tested if the origins of SNe Ia are recognized.
Knowledge of their nature is also of importance for  understanding the  
metallicity evolution and star-formation  history of galaxies.
Yet, despite their relevance, no concensus on the  
nature of their progenitor systems has been reached.

While there is broad agreement that the disintegration of a  white dwarf  
(WD)  in a thermonuclear explosion constitutes the  supernova event  
itself,  there are two main classes of competing models for the  
events  that lead to the explosion. In the single-degenerate   
scenario,  the exploding WD accretes from a non-degenerate stellar  
companion \citep{whelan73,nomoto82}, which is expected to survive and  
be potentially identifiable. In the double-degenerate scenario, the  
donor star is  also a  WD.  The most commonly discussed progenitor  
system involves the coalescence of two CO WDs  
\citep{iben84,webbink84}, which after explosion should leave no  
remnant.   There has been no conclusive proof to date that either 
scenario can lead to normal SNe Ia, nor  
has the evidence that the SN Ia rate is different for different  
stellar populations led to any firm conclusions. Therefore, any new  
observational or theoretical constraint on the progenitor systems is  
of great value.

  Here we present an alternative evolutionary scenario to produce a  
SNe Ia, which is not restricted to mass transfer in gravitationally
bound double stellar  systems. This new paradigm considers white dwarfs 
that reside within dense stellar systems where the stars  are
sufficiently  close to each other  to make collisions quite likely.  
The resultant shock compression could then lead to densities which
 exceed the threshold  for pycnonuclear  
reactions so that thermonuclear runaway ensues. Understanding the   
feasibility of this channel for producing successful thermonuclear  
explosions as well as exploring the observational manifestations of  
such phenomena are the main purpose of this {\it Letter}. The layout  
is as follows. A concise summary of the numerical methods and the  
initial models is given in \S~2. We describe the detailed hydrodynamic 
simulations in \S~3, while the resulting broadband lightcurves and spectra 
together with a discussion of the relevance of this new progenitor 
channel to upcoming supernova surveys are presented in \S~4.

\section{Numerical Schemes and Initial Models}
As two stars approach each other to within a few stellar radii, their mutual gravitational
interactions lead to the development of large-scale tidal distortions that substantially alter
their global structures. If the trajectories of the stars bring them so close to each other that they experience a
collision, the response of the stellar material to the impact is critical in understanding the
future evolution of the system. As such, it is no longer appropriate to treat the stars as point
masses, and a hydrodynamical description of the encounter becomes necessary.
The outcome
of a collision between two white dwarfs depends in an essential way 
on several factors: their masses and nuclear compositions, their 
relative speed, and the distance of closest approach.

To study this problem, we use two complementary approaches: the Eulerian,
adaptive-mesh hydrodynamics code FLASH  \citep{fryxell00}, and a Lagrangian 
hydrodynamics code \citep{rosswog08b,rosswog09a} based on the Smoothed Particle 
Hydrodynamics (SPH) method \citep{benz90a,monaghan05,rosswog09b}. Both codes 
incorporate the Helmholtz equation of state \citep{timmes00a} and similar, small nuclear 
reaction networks tuned to correctly reproduce nuclear energy release \citep{hix98,timmes99,timmes00b},
but they differ in their treatment of hydrodynamics and gravity. Using the same 
stellar models and impact conditions, this approach not  only provides a classic 
code verification, but in particular allows to gain unique insight into the physics of 
white dwarf encounters.

Some of the questions at the forefront of our attention are the  
effects of the initial nuclear composition and masses of the white 
dwarfs as well as the impact conditions. We have performed a large 
set of three-dimensional calculations. A detailed account of all 
models will be given elsewhere, here our focus is on central 
collisions. A summary of the performed calculations is given in Table 
\ref{table1}. The stars with 0.4 \Msun are instantiated as pure He, 
those with larger masses as a homogeneous mixture of 50\% C and 50\% O. All stars are initially 
cold ($10^4$ K), placed at a mutual distance of three times their 
combined unperturbed stellar radii and with the relative free-fall 
velocities of the corresponding point mass values.\\
\begin{table*}
 \centering
   
  \caption{Masses in \msun, initial separation $a_0$ in $R_1+R_2$, 
           densities and energies in cgs-units, "res." refers to the particle number 
           for SPH calculations, and to maximum linear resolution in  cm for FLASH.}
  \begin{tabular}{@{}llrclllllll@{}}
  \hline
   run    &    masses & $a_0$& res. & $\log(\rho_{\rm max})$ & $T_{{\rm max},9}$ & 
$\log(E_{\rm nuc})$ & $m_{\rm esc}$  & remnant\\
 \hline
   &&&&& {\bf SPH} &&&&&\\
A & 0.2, 0.2 &  5 & $2.0\times 10^{5}$ & 5.97 & 2.5 & $48.53$ & 0.044 & hot, He-WD \\
B & 0.4, 0.4 &  3 & $2.0\times 10^{6}$ & 7.16 & 4.0 & 51.19   & 0.80  & none\\
C & 0.5, 0.5 &  3 & $1.0\times 10^{6}$ & 7.20 & 4.8 & 50.00   & 0.21  & CO-WD in Ne-Mg-Si cloud\\
D & 0.6, 0.6 &  3 & $2.0\times 10^{6}$ & 7.92 & 8.9 & 51.21   & 1.20  & none\\
E & 0.4, 0.9 &  3 & $2.5\times 10^{6}$ & 7.56 & 3.4 & 50.75   & 0.40  & CO-WD in He-Si cloud \\
F & 0.6, 0.9 &  3 & $2.0\times 10^{6}$ & 8.40 & 7.9 & 50.61   & 0.30  & CO-WD in C-O-Si-Fe cloud\\
G & 0.9, 0.9 &  3 & $1.0\times 10^{6}$ & 7.55 & 6.3 & 51.41   & 1.80  & none\\
   &&&&& {\bf FLASH} &&&&&\\
H & 0.6, 0.6 &  3 & $4.9 \times 10^6$ & 7.47 & 5.5 & 51.11   & 1.20  & none\\
\label{table1}
\end{tabular}
\end{table*}

\section{Shock-Triggered Thermonuclear Explosions from White Dwarf  
Impacts}

\begin{figure*}
\centerline{\includegraphics[width=17cm,angle=0]{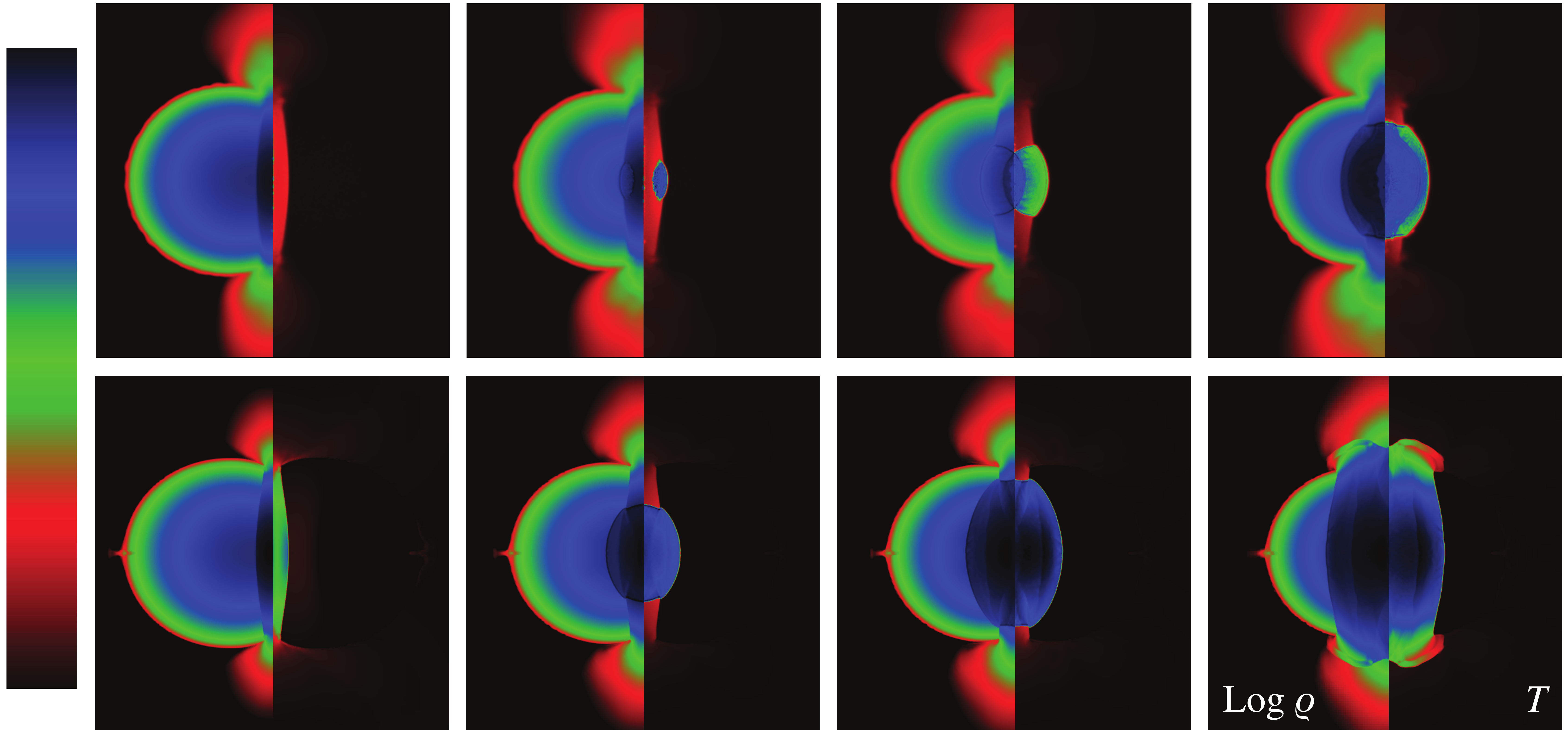}}
\caption{Comparison of density and temperature evolution of the central collision of
two 0.6 \msun CO white dwarfs. The upper two rows are the SPH-results, the lower two
are produced by FLASH. The shown box length is $3 \times 10^9$ cm, limiting values colorbar (left to right):
         $\log(\rho)_{\rm max}= [7.12,7.05,6.91,6.84,6.70,6.64]$, $T_{\rm 9, max}=[1.49,4.79,3.99,3.65,3.39,3.24]$, $\log(\rho)_{\rm min}=2$ and $T_{\rm 9, min}= 0$ everywhere.}
\label{fig1}
\end{figure*}

The relative velocity at contact is entirely dominated by the mutual gravitational 
attraction, i.e. it is much larger than typical GC velocity dispersions, 
$\sigma_{\rm GC} \approx 10$ km/s,
$v_{\rm rel}= 4000 \; {\rm km/s} \; (M_{\rm tot}/1.2$ \msun)$^{1/2} (2 \times 10^9 {\rm cm}/(R_1+R_2))^{1/2} > c_{\rm s}$.
The sound velocity $c_{\rm s}$ in the core of a 0.6 \msun WD is about 2600 km/s, and thus shocks are a natural result of a WD collision.
Figure \ref{fig1} shows the thermodynamic evolution of the most common combination
of masses, 2 x 0.6 \msun. The snapshots of density and temperature in the upper two rows were
obtained with SPH, the lower two are the result of FLASH (19-isotope network, minimum linear
resolution $5\times 10^6$ cm). The details
of the collision differ slightly in the two simulation environments, which is evident in the shock geometries. However, the overall behavior is similar: shortly 
after contact, a discus-shaped, shock-heated region forms in which nuclear processing occurs. 
Since the ignition site is not coincident with the original central density peak of either WD, the shocks more quickly propagate through the shallow density gradient that is perpendicular to the direction of infall. As a result, the hot processed material first breaks out through a ring which lies in a plane that is parallel to the collisional plane. It is only when the rear of the 
stars have passed through the shock fronts that a significant overall expansion can set in. 
Apart from hydrodynamics and gravity, the codes also differ in the used reaction networks, the SPH code is coupled
to a 7-species network \citep{hix98} while the FLASH run uses a 19-isotope network \cite{timmes99}.
As a test of the energy generation accuracy we have used both networks along 1000 thermodynamic trajectories 
extracted from the SPH simulation. Maximum deviations in the resulting energy generation were 15\%, while 95\% of  the trajectories agreed to better than 5\%. The mass fractions reported in this paper are all post-processed
or direct results of the 19-isotope network. While the networks could be partly responsible for the shock 
structure differences the latter
may also be influenced by the local resolution and details how burning directly in the shock
is suppressed in both codes. A further difference between both runs is the spike-like feature at the rear side of the star in the
FLASH simulation, this is an artifact of the rapid advection across the grid. Despite these differences 
our main conclusions are robust: the shocks trigger an explosion producing a substantial amount of 
$^{56}$Ni (0.32 \Msun for SPH, 0.16 \Msun for FLASH).

\begin{figure*}
\centerline{
\includegraphics[width=8.5cm,angle=0]{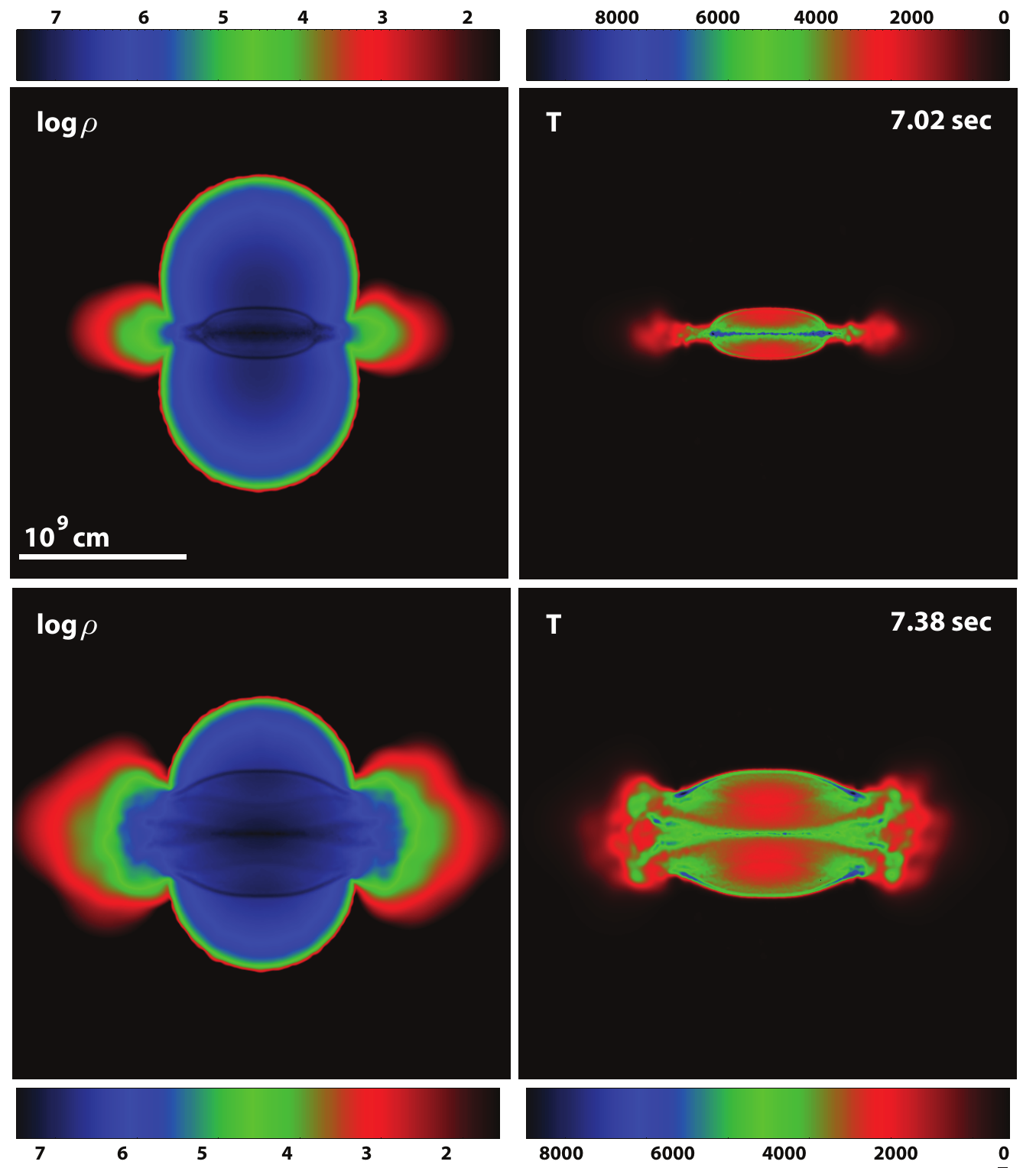}
\includegraphics[width=8.5cm,angle=0]{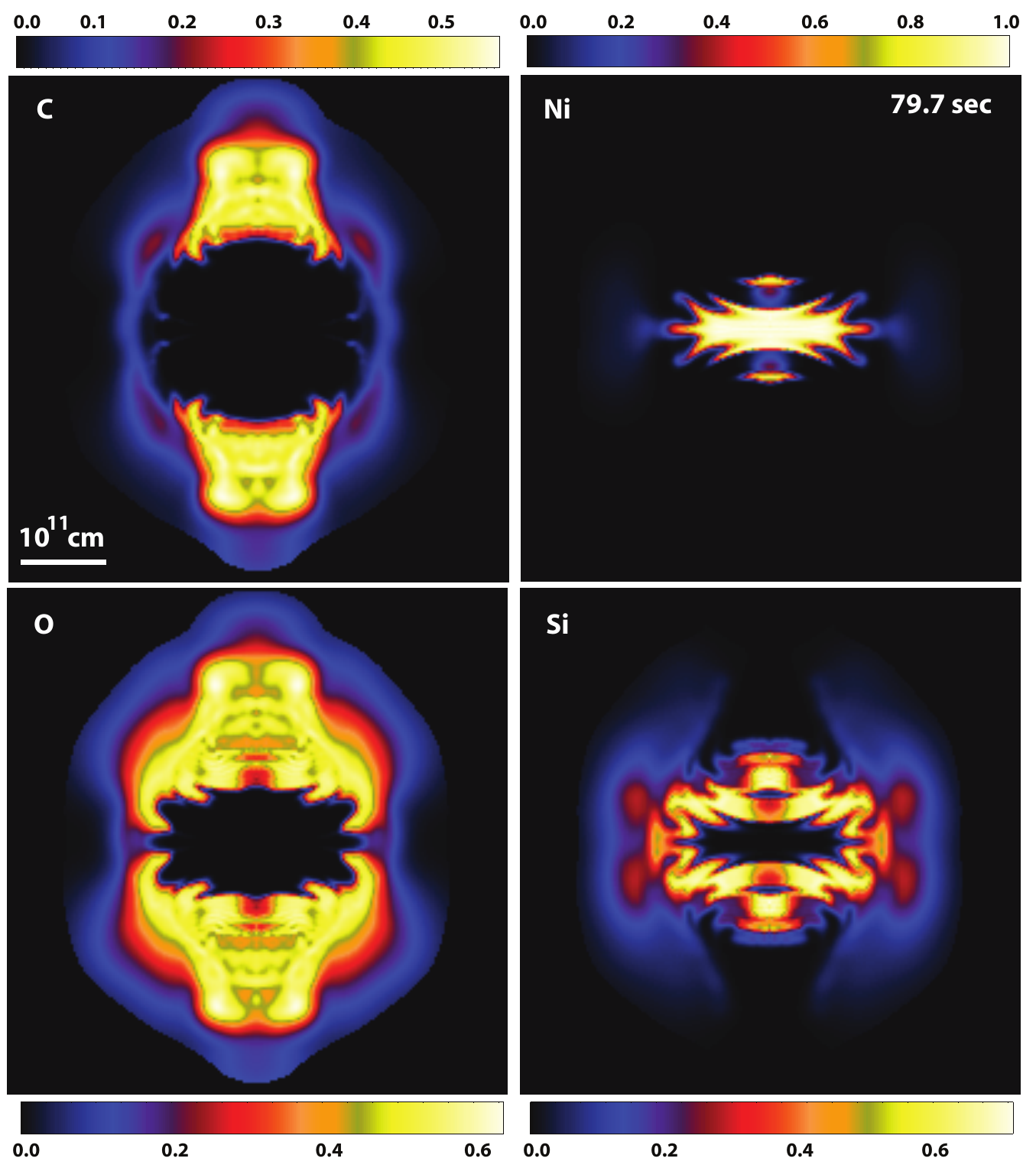}
}
\caption{Collision between two 0.9 \msun white dwarfs: density and temperature (left), 
nuclear mass fractions (right). Note the different scales in both panels.}
\label{fig2}
\end{figure*}

Mass-segregated environments may favour encounters of more massive WDs, though. We therefore show in 
Fig.~\ref{fig2} the outcome of a 2 x 0.9 \Msun 
collision (run G; left: density and temperature, right: mass fractions). This event yields 0.66 \Msun 
of $^{56}$Ni, comparable to a typical SNIa.

The topic of white dwarf collisions has been pioneered by \cite{benz89}. In their work 
they modeled the white dwarfs with an equation of state containing contributions 
from  degenerate, relativistic electrons, from a Boltzmann gas of nuclei and  from photons. 
Moreover, they coupled their smooth 
particle hydrodynamics code to  a 14-isotope network. At that time their calculations were 
restricted to 5000 SPH particles resulting in a moderate numerical resolution. Although in some cases 
substantial nuclear burning took place,  none of their models resulted in a complete disintegration 
of the white dwarf pairs. For comparison, we performed a test run of a headon collision 
between two 0.6 \Msun WDs using 5000 SPH particles in total, similar to their run 1. Consistent 
with their work, we find a surviving remnant and only 0.30 \Msun of expelled material 
(0.09 \Msun in their work). The degraded numerical resolution results in  a one order of 
magnitude reduction of the released nuclear energy  compared to our run D. The remaining 
differences between our test run and the results of Benz et al. are mainly due to the 
different equation of state, but they may to some extent also reflect the differences in 
the networks and the advances in the SPH method. Our overall results are similar 
to those of \cite{raskin09} which were submitted while our paper was under review.

\section{Discussion}

\subsection{Lightcurves and Spectra}
Some white dwarf collisions should produce luminous light curves
powered by the decay of radioactive \Nifs\ synthesized in the
explosion.  To predict the observable emission, we post-processed
select models using the radiative transfer code SEDONA
\citep{kasen06}. This code is 3-dimensional, time-dependent,
multi-wavelength and includes a detailed treatment of the physics of
\Nifs\ decay and bound-bound line opacity.  As initial conditions, we
used the SPH results with post-processed abundances for the 2 x 
0.6~\Msun\ (run D) and 2 x 0.9~\msun\
(run G) collisions, at a simulation time late enough that the ejected
material had reached the free, homologous phase of expansion.  Given the axial
symmetry of head-on encounters, the results where azimuthally averaged onto a cylindrical grid.

\begin{figure*}
\centerline{\includegraphics[width=17cm,angle=0]{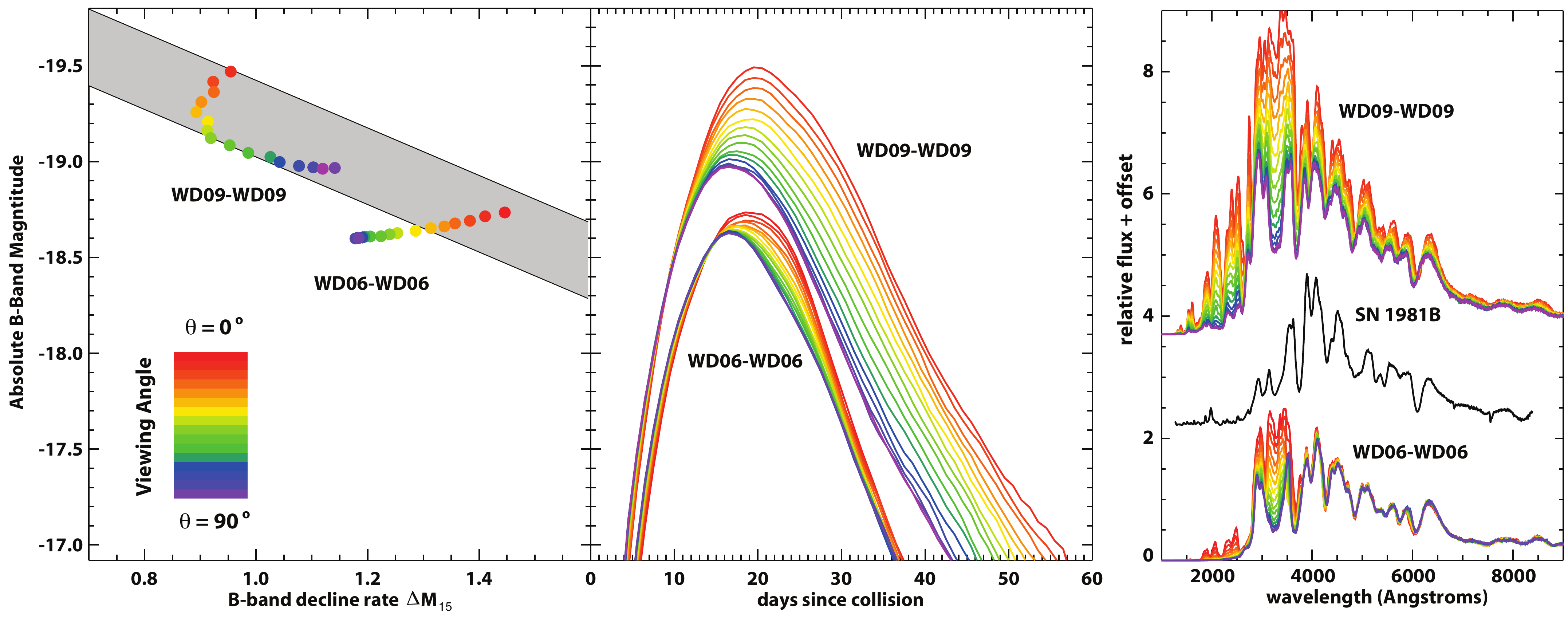}}
\caption{Radiative transfer calculations of the light curves and
spectra resulting from central collisions of 2 x 0.6~\msun\ (run D)
and 2 x 0.9~\msun\ (run G) white dwarf pairs.  The synthetic B-band
light curves (central panel) closely resemble those of normal Type~Ia
supernovae.  The peak brightness and decline rate of the light curves
vary somewhat with the viewing angle (left panel), but are broadly
consistent with the slope and spread of the observed Phillips relation
(grey shaded band).  The maximum light spectra (right panel) closely
resemble that of the normal Type~Ia supernova SN~1981B.} \label{fig3}
\end{figure*}

In Figure~\ref{fig3} (right panel) we show synthetic spectra of the
models, computed at the peak of the light curve ($\sim 20$~days after
collision).  The model spectra closely resemble those of normal
Type~Ia supernovae, with broad P-Cygni line features due to Si~II,
S~II, and Ca~II.  This outcome is not surprising given that the ejecta
compositional stratification (Figure~\ref{fig1}) is very similar to
that of standard SNe~Ia models, with an outer layer of intermediate
mass elements and an inner core of iron group elements.  The asymmetry
of the ejected debris introduces some variation of the spectrum with
viewing angle, most prominently in the ultraviolet where the radiation
transport is most sensitive to line blanketing opacity.  The velocity
of the supernova photosphere, as measured form the Doppler shift of
the line absorption minima, is $13000-16000$~km~s$^{-1}$, similar to,
though slightly higher than that observed in average SNe~Ia.

The light curves of the white dwarf collisions (Figure~\ref{fig3},
center panel) also resemble those of Type~Ia
supernovae.  As expected, the models which produced more \Nifs\ have
brighter light curves which decline more slowly.  The calculated peak
magnitudes and B-band decline rates lie within the range of typical
SNe~Ia, and are consistent with the slope and normalization of the
observed Phillips relation (Figure~\ref{fig3}, left panel).  This result is not
totally unexpected, given that the principle parameter underlying the
Phillips relation is the \Nifs\ mass, which influences both the
supernova luminosity and the ejecta opacity \citep[][and references
therein]{kasen07}.  On the other hand, the light curve width is also
sensitive to the total ejected mass.  In collision models, unlike
most standard SN~Ia scenarios, this value is not constrained to be the
Chandrasekhar mass. Thus, although the particular models studied here
roughly obey the Phillips relation, in detail white dwarf collisions
could show small but systematic deviations.

\subsection{Diversity}
The explosion mechanism reported here is not tied to a particular mass scale 
and therefore allows for considerably more diversity. As mentioned above,
collisions between white dwarfs provide a pathway to ignite CO white dwarfs
that completely disintegrate the WD pair. In contrast, for low-mass collisions
(run A) or for collisions between a CO and a He WD a  remnant remains, which, in the
latter case, produces an identifiable outcome: a hot, high-speed ($\sim 1000$ km/s)
CO WD engulfed by a cloud of intermediate mass elements.

The mechanism discussed here is found to work for the  collision of two 0.4 \Msun He
WDs, but does not lead to an explosion in the case of 2 x 0.5 \msun. This is mainly the result
of the different available fuel -- helium burning releases more energy on the way to iron group elements
($\epsilon_{\rm He \rightarrow Fe}= 1.73$ MeV/nucleon, $\epsilon_{\rm
CO \rightarrow Fe}= 0.98$ MeV/nucleon). Due to the WD mass function,
collisions  between 0.6 \Msun WDs are expected to dominate  unless they occur in a strongly
mass-segregated environments where more massive WDs would then  be preferred.
On average, however, this SN Ia channel will preferably ignite lighter WDs than standard
channels
\citep{hillebrandt00,podsiadlowski08} and, as a result, the nucleosynthetic yields
should be less neutron-rich due to the slower $Y_e$-evolution via electron captures.

\subsection{Detection Prospects}
In order to critically evaluate the outcome of these ideas, the  
frequency of such
encounters must be addressed.
For a core-collapsed GC, the high densities in the core completely
dominate the collision rate. We assume the white dwarfs
to be distributed homogeneously within a spherical core of
radius $r_{\rm c}$.  We further assume that the total number density
$n_{\rm wd}$ and stellar velocity dispersion $\sigma_{\rm c}$ are
constant within $r_{\rm c}$. Together with the dominating
gravitational focusing, this means we can approximate the
total collision rate as: $\nu_{\rm col} = 20 \;{\rm
Gyr}^{-1}\;f_1\,f_2(n_{\rm c}/3 \times 10^6\;{\rm pc}^{-3})^2 (r_{\rm
c}/0.1\;{\rm pc})^3(\sigma_{\rm c}/ 10\;{\rm km\,s}^{-1})^{-1}$
$([m_1+m_2]/1\;{\rm M}_\odot)(r_{\rm col}/5 \times 10^3\;{\rm km})$,
where
$f_i\le1$ is the fractional number of stars of type $i$ within $r_{\rm
c}$ and we use  the properties of the well-studied, proto-typical core-
collapsed
GC M15 \citep{vandenBosch06}. Fokker-Planck model fits to M15  
predict the presence of a significant population of WDs
with $M>0.7M_\odot$, so that  $f_i>0.5$. 
By setting the distance of closest approach to the sum
  of the radii of the two WDs, $r_{\rm col}=r_1+r_2$, an 
  estimate of the collision rate 
 $R_{\rm col}\approx 10^2 f_{\rm cc} f_1\,f_2$ yr$^{-1}$ Gpc $^{-3}$ , 
 can be obtained by multiplying $\nu_{\rm col}$ (per GC) with 
 the average GC space density of 
 $n_{\rm gc}=4.2\;{\rm Mpc^{-3}}$ \citep{brodie06}. Here  
 $n_{\rm gc}$ is derived by combining the number of GCs per 
 galaxy with the galaxy luminosity density distribution and 
 $f_{\rm cc}$ is the fraction of core-collapsed clusters.
This is most likely an underestimation since,  
for example, the effect of binaries in GCs can increase $\nu_{\rm
col}$ by a moderate factor ($\sim 2$, J. Fregeau, 
private communication). Moreover, if M\,15 formed at
higher initial concentration, it might have been in (or around)
deepest core collapse for a longer time, significantly increasing the
(average) $\nu_{\rm col}$. Numerical experiments for the dominant
2 x 0.6  \Msun case indicate that about 20\% of the collisions may
result in explosions.

Although these rates are still subject to significant
uncertainties such as whether other GCs follow a similar
core-collapse evolution, they indicate that white dwarf collisions in
their dense cores are not unlikely and can contribute with a modest
fraction to the SNe Ia population, whose event rates are estimated
to be of order a few  $10^4$ yr$^{-1}$ Gpc$^{-3}$ \citep{cappellaro99}.
In addition, a number of collisions are also expected from  ultra-
compact dwarf galaxies \citep{hilker99,drinkwater00,drinkwater03}, the
hypercompact stellar systems that form when supermassive black holes
are ejected from galactic centres
by the action gravitational wave recoil \citep{merritt09} and from more
"typical" galactic centers.

The transient sky at faint magnitudes is poorly known, but there are
major efforts under
way that would increase the discovery rate for type Ia supernovae from
a few thousands
to about hundreds of thousands per year.
While the estimates given above are much too low to explain the bulk of
the SNe Ia
population, they may be frequent enough to provide upcoming supernova
surveys with hundreds of collision-induced SNe Ia per year.

\acknowledgments We thank Holger Baumgardt, Lars Bildsten, John Fregeau, Ken 
Shen and Glenn van de Ven for very useful discussions. The simulations presented 
in this paper were performed on the JUMP computer of the H\"ochstleistungsrechenzentrum 
J\"ulich and the  Pleiades computer of UCSC. We acknowledge support from
DFG grant RO 3399 (S.R.), the DOE Program for
SciDAC DE-FC02-01ER41176 (D.K and E.R.) and The David and Lucile  
Packard Foundation (J.G. and E.R.). Support for DNK was provided by NASA through 
Hubble fellowship grant
\#HST-HF-01208.01-A awarded by the Space Telescope Science Institute,
which is operated by the Association of Universities for Research in
Astronomy, Inc., for NASA, under contract NAS 5-26555.

\hyphenation{Post-Script Sprin-ger}

\clearpage

\clearpage

\clearpage

\end{document}